# Surpassing the Ambient Packing Limit of Energetic Crystals: Squeezing Effect of Molecular Level "Net-fishing"

Qi-Long Yan*[1], Zhi-Jian Yang[2], Guo-Qiang He[1]*, Wei He[1], Jie-Yao Lv[1], Shi Huang[2], Jian Chen[1], Shu-Wen Chen[1], Pei-Jin Liu[1], Qing-Hua Zhang[2], Fu-De Nie[2]

1. Science and Technology on Combustion, Internal Flow and Thermo-structure Laboratory, Northwestern Polytechnical University, Xi'an 710072, China;
2. Institute of Chemical Materials, CAEP, Mianyang, 621900, China.

**Abstract:** High energy density is always a key goal for developments of energy storage or energetic materials (EMs). Except exploring novel EMs with high chemical energy, it is also desirable if the traditional EMs could be assembled at a higher density. However, it is very difficult to surpass their theoretical maximum molecular packing density under ambient conditions, even though a higher density could be achieved under ultra-high pressure (Gigapascals). Such solid-state phase changes are reversible, and hence this high density packing is not able to maintain under ambient conditions. Alternatively, in this research, we demonstrated a molecular level compression effect by stacking of 2-D TAGP, resulting in a higher density packing of the HMX molecules with changed conformation. The HMX crystal formed under compression in the solvent has a unit cell parameter very close to the reported one observed under pressure of 0.2 GPa. It shows that the compressed HMX molecules are trapped in the TAGP layers, resulting in a higher density (e.g. 2.13 g cm$^{-3}$) and also higher heat of formation. The resulted HMX crystals are free of defects, and unlike the pristine HMX, no polymorphic transition and melting point were observed upon heating. Experiments and relevant calculations show that the best resulted hybrid HMX crystal has a detonation velocity of 10.40 km s$^{-1}$ and pressure of 53.9 GPa, respectively. Its ground specific impulse reaches about 292 s, even much better than CL-20, making it a promising propellant component for future space explorations.

It is essential to develop energy storage or energetic materials (EMs) with higher energy densities. In recently years, more and more attention has been paid to this field, due to high demands in deep-space exploration, aerospace and defense technologies. Typical novel highly EMs include energetic ionic liquids (1), metastable intermixed composites (MICs) (2), novel insensitive nitro-compounds (3), full nitrogen compounds (4,5), polynitrogen compounds (6), boron-nitrogen based EMs (7), as well as carbon nanomaterials based energetic nanocomposites (8). However, most of these newly developed high EMs are still not qualified to replace currently used EMs due to various problems including chemical incompatibility, thermal instability, high sensitivity, as well as high cost. Another group of fancy disruptive EMs is so-called extended solid, and the typical representatives are metallic hydrogen (9) and poly-CO (10). The solid molecular hydrogen under extremely high pressure (e.g. >495 gigapascals) would become metallic with reflectivity over 0.9, which is considered as a room-temperature superconductor and a very promising as a rocket fuel if it could be stabilized at mild conditions. However, these materials are only stable under extreme conditions. Theoretically, they are revolutionary in structure and performances, but still much more effort has to be made before their practical applications.

In spite of great progresses in development of novel EMs, the widely used energetic ingredients in formulations are dominant by 1,3,5-trinitro-1,3,5-triazinane (RDX), 1,3,5-triamino-2,4,6-trinitrobenzene (TATB), 1,3,5,7-tetranitro-1,3,5,7-tetrazocane (HMX), hexanitrohexaazaisowurtzitane (CL-20) and aluminum (Al), (11). For enhanced performances, these conventional EMs materials have to be modified by new processing technologies. Much research work has been done on this issue in the past decades, where successfully strategies include particle polishing (13), coating (14), co-crystallization (15), and doping of 2D materials such as graphene (16). Alternatively, if the abovementioned energetic molecules or fuels could be packed at a higher density during modification, it would be highly desirable, since the performances of these materials are largely dependent on their density. However, it is very difficult to surpass their theoretical maximum density under ambient conditions, even though a higher molecular density packing could be achieved due to conformation changes of the component molecules under ultra-high pressure (Gigapascals, usually by using a diamond anvil) (17). It has been demonstrated that such solid-state phase changes are reversible, and hence this high density packing is not able to maintain at ambient conditions.

---

* Corresponding authors: Q.-L. Yan, qilongyan@nwpu.edu.cn; G.-Q He, gqhe@nwpu.edu.cn;

Aiming to achieve an irreversible high density packing, a molecular level *in-situ* compression of energetic molecules during their crystallization in the solutions would be an appropriate choice. However, there are seldom chemical reactors can reach pressurization over a gigapascal due to limited sealing technology and material strength. Interestingly, we found that a molecular level compression effect could be achieved between layers of 2-D materials during their stacking under high temperature (~150 ˚C), resulting in large conformation change of energetic molecules trapped between these layers. This 2-D material was actually recently reported, which has large hollow defects on its layered structure (18). It is an analogue of graphitic carbon nitrite (g-$C_3N_4$) named as TAGP.

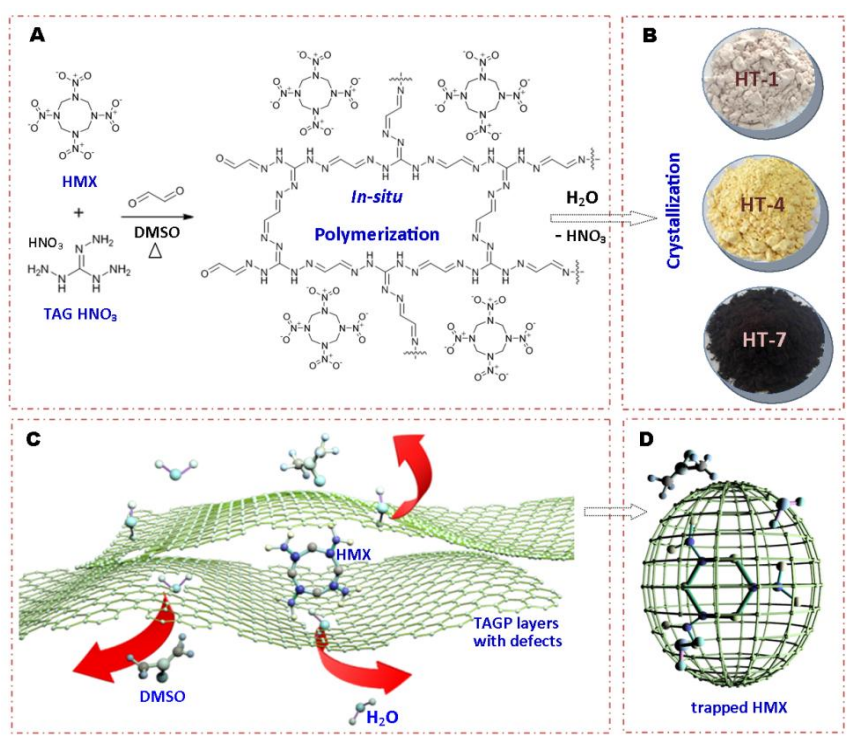

**Fig. 1. The reaction mechanisms and proposed concept of the "net-fishing":** (**A**), the chemical reaction mechanism of triaminoguanidine nitrate with glyoxal in presence of dissolved β-HMX in DMSO. (**B**) Digital photos of typical as-prepared products: crystalline powders of HT-1, HT-4 and HT-7 (DMSO/$H_2O$ solvent/unti-solvent system). (**C**) Scheme of the packing and squeezing process of the DMSO molecules from HMX aggregations by 2-D TAGP stacked layers with hollow defects, which is therefore vividly so-called a "net-fishing" process. (**D**) The trapped HMX molecules crystalizes in enclosed TAGP layers as the dopants, as shown in Fig. 2H.

To symbolically depict this compression and the following molecular packing process, herein the term "net-fishing" has been selected for a vivid description. As a demonstration, high energetic nitramine HMX molecules were used for this research, and the corresponding reaction scheme is presented in Fig. 1A. The dragging force of the "net-fishing" process is the strong interaction of π-π stacking among the *in-situ* generated TAGP networks in concentrated DMSO solution, and the preconditions for squeezing the DMSO molecules out is the absorption force of water as the unti-solvent and the large hollow defects presented on the TAGP layers which act as the "mesh". Such stacked "mesh" has a size of one or two nanometers, so that it is difficult for the large HMX molecules to travel through as the trapped "big fishes", whereas the solvent DMSO molecules were squeezed out (Fig.1C). Also, the hydrogen bonds between the –NO$_2$ group and –NH– on the TAGP skeleton is another force that preventing HMX molecules going out. It has been found that, various doped/modified HMX crystals could be obtained under different experimental conditions. This is a very facile and flexible method combining *in-situ* cross-linking and co-crystallization techniques, where the insensitive high-nitrogen 2D TAGP networks can be integrated or doped in HMX crystals. The TAGP layers played a dual role as both compressing "anvils" and dopants, so that various hybrid energetic crystals with different structure and performances were obtained depending on the experimental conditions (Details of the experiments and nomenclature of the typical products are presented in Supporting Document). Typical selected products with different colors are shown in Figure 1B and fig. S1.

As shown in fig. S1, the cream white industrial grade β-HMX was transformed into various colors, including yellows, grey, even dark brown and black due to change of conformation and a minor doping of TAGP networks. In order to compare the crystal morphologies of these products, the SEM images (Figs. 2A-G, fig.S2) coupled with elemental analyses by EDS (Figs. 2p$_1$-p$_4$, fig.S3), as well as TEM images (Figs. 2H-L) were obtained. As shown in these figures, it is clear that the doped HMX crystals obtained under different conditions differs much in morphology. In general, the shapes of those HMX crystals can be simply divided into three categories: polyhedron (Figs. 2A and 2E), stick (Figs. 2B and 2C), and flake (Figs. 2D, F, and G). It is clear that the slow growth of crystal H-1 and H-4 would result in much different morphology than their counterparts obtained by fast crystallization. However, in the case of HT-7, both fast and slow crystallization lead to the same flake structure, where the only difference is the dimensions (aggregations of tiny flakes *vs.* large plates in hundreds of microns). It means at higher temperature, a larger TAGP network could be formed with less defects, and under the guidance of these layers, the crystal could be inclined to grow in 2-dimensional.

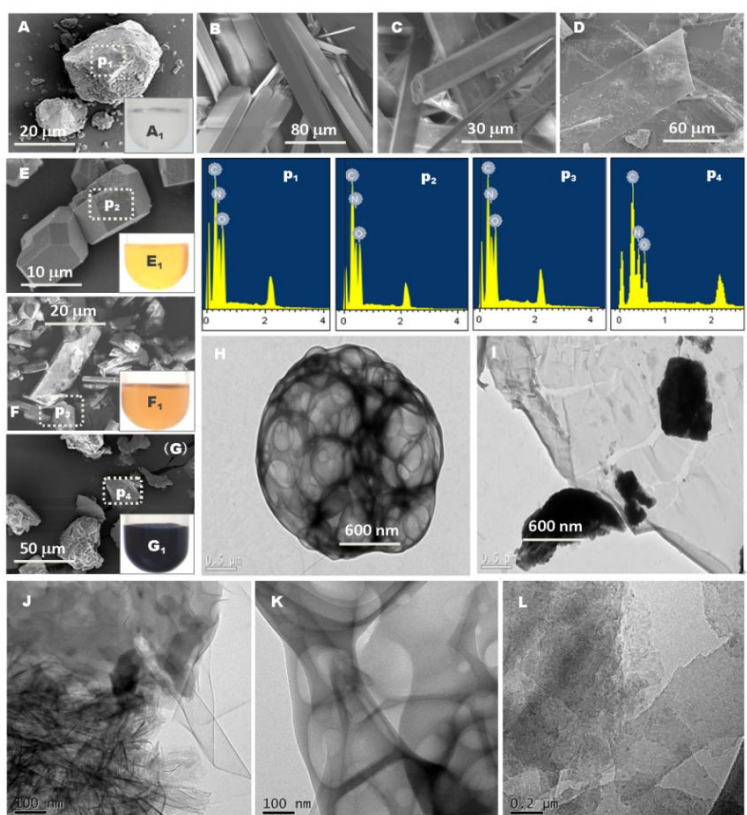

**Fig. 2. The SEM/EDS and TEM images.** (**A**), the SEM images of pristine β-HMX as the starting material. (**B**), (**C**) and (**D**), SEM images of HT-1, HT-4 and HT-7 crystals obtained by slow crystallization process (DMSO/H$_2$O solvent/unti-solvent system, over one month's water diffusing and dissolving in DMSO at room temperature). (**E**),

(**F**) and (**G**), SEM images of HT-1, HT-4 and HT-7 crystals obtained by slow crystallization process (DMSO/H$_2$O system, water adding in 10 minutes). (**H**) and (**I**), TEM images of nanomaterials that disperse in the DMSO/H$_2$O solutions after crystallization process of HT-4 and HT-7, respectively. (**J**), (**K**) and (**L**), TEM images of the separated dopants by slow half evaporation of saturated solution of HT-1, HT-4 and HT-7 crystals using acetone/ethyl acetate with 1/1 volume ratio. (**A$_1$**), (**E$_1$**), (**F$_1$**) and (**G$_1$**), the appearance of the DMSO saturated solution of β-HMX, HT-1, HT-4 and HT-7 at room temperature, respectively. (**P$_1$-P$_1$**), EDS spectra of the marked areas in the corresponding SEM images.

As a common fact, the nitramine crystals are very sensitive to the electron beam used in SEM and TEM techniques. It is clear from the presented images of SEM (Fig. 2F) and TEM (Fig. 2I) that some black spots are shown on the surfaces of the involved crystals after short-exposure to the electron beam even at lower voltage e.g. 5 kV, where the HT-0, HT-1, HT-2 and HT-4 are even more sensitive than the others. The nontransparent blocks shown in Figs. 2I and J are the carbonized remnant HMX crystals after high voltage electron beam shooting (19). Typically, there are two types of 2D TAGP materials formed in the mother DMSO solution. One is curly and round-packed layers of TAGP with large hollow defects, look like a fishing-net (Figure 2H). Another one is the smooth transparent graphene-like film without any defects. In the presence of highly concentrated HMX molecules, a part of the original defected TAGP skeleton could be repaired. As a demonstration, after dissolving of the crystals of HT-1, HY-4 and HT-7 in a mixture solvent of acetone and ethyl acetate, the doped TAGP materials could be separated by simply evaporate about 1/3 of the saturated solution. They are very different in morphology: the TAGP inside HT-1 is a mixture of nano-fibers and thin films (Figure 2J), and it is hollowed thin films in HT-4 (Figure 2K), whereas defect free 2D layers are presented in HT-7 crystals (Figure 2L).

HMX is a flexible eight-membered ring compound and it can exist in different conformations that lead to different crystal structures having different thermal stability, sensitivity, and reactivity. These are called α-, β-, γ- and δ- polymorphs (20). The β-HMX is most stable and is widely used for various applications. At the beginning, we believe that these crystals are hybrid crystalline materials (multi-phases, dopants with various molecular weights), and it could be very difficult to analyze their structures by using single crystal X-ray diffraction. Therefore, to clarify their structure discrepancies, except for MS and NMR spectra (Fig. 3A, and figs. S6, S7 and S12), various popular techniques used in material science have been attempted, including FTIR (fig. S4), Raman (fig. S5), XPS (Figs. 3B, 3C, S10 and S11) and powder XRD (Fig. 3E, figs. S8 and S9). It is clear from MS, FTIR and NMR spectra that these techniques would not show us any difference between the pristine HMX and the modified/doped HMX crystals, but one thing can be confirmed is that the chemical bonding structure does not change during modification. The Raman spectra is greatly dependent on the capability of the light scattering for the samples, therefore the crystals that are in deep color (e.g. dark brown HT-3 and black HT-7, see in fig. S1) would have less peaks or broader peaks. As reported in the literature, HMX displays a fingerprint-like quality of β-HMX in the region of 200–600 cm$^{-1}$, whereas peaks in the 800–960 cm$^{-1}$ and 1000 to 1500 cm$^{-1}$ region are attributed to the ring-stretching vibration and symmetric vibrational stretching of the NO$_2$ and N–NO$_2$ groups of HMX molecules, respectively. Only HT-7 does not show these typical peaks due to very weak light scattering, which contains the highest mass ratio of TAGP dopant among all prepared samples (about 24.67%), whereas the HT-4 contains TAGP of about 1.96 wt% determined by GPC method (fig. S13).

The deconvolution of XPS spectra (figs. S10, S11) and elemental analyses (see in figs. S3, S4 and Table S1) show successful doping of TAGP materials inside HMX crystals. The presence of the C-C and N-H bonds, with binding energies of about 283.4 eV and 398.3 eV, respectively, confirmed the presence of the –[NH–N–C(=N–N–)–N–NH–C–C]$_n$– networks (Fig. 1A) inside the modified crystals of HMX. the Elements recorded from the combustion experiments show that after doping, there is about 0.06-0.13 wt.% increase in C content and 0.09-0.20 wt.% in O content, whereas the N content decreases by 0.22-0.24 wt.% with H almost unchanged. In order for better comparison, these changes are based on the measured elemental content of HMX as the starting reference material. The measured O content of HMX is about 1.43 wt.% higher than theory, while the other elements are lower. Since the dopants TAGP are different in size (extent of crosslinking) and structure as shown by TEM due to different experimental conditions. Therefore, we could not simply use a formula to represent these TAGP herein. For a better comparison, the experimentally measured formula with normalized number of carbon atoms for the modified HMX and raw β-HMX crystals are presented in Table S1. However, such a small change in elements due to TAGP doping would result in large difference in crystal structure and density. Initial powder X-ray spectra show that at least HT-1, HT-2, HT-4 and HT-7 does not belongs to any form of polymorphs of existing HMX or their mixtures, showing extra peaks (e.g. 18.6˚, 34.1˚) or shifted peaks at various diffraction angles (figs. S8, S9). There is also no obvious TAGP diffraction peaks due to a minor content of this 2D material in these modified crystals. The calculated space groups from PXRD patterns for these modified HMX are quite different from the existing polymorphs of HMX, except HT-0, HT-5 and HT-8 (Table S2), which shows almost identical PXRD pattern as β-HMX (fig. S8). It means that the TAGP material does not go inside the crystal lattice of β-HMX, resulting in a decreased density. The modified crystals with different crystal structure have higher density than pristine β-HMX, where HT-7 has the highest density of 2.13 g cm$^{-3}$.

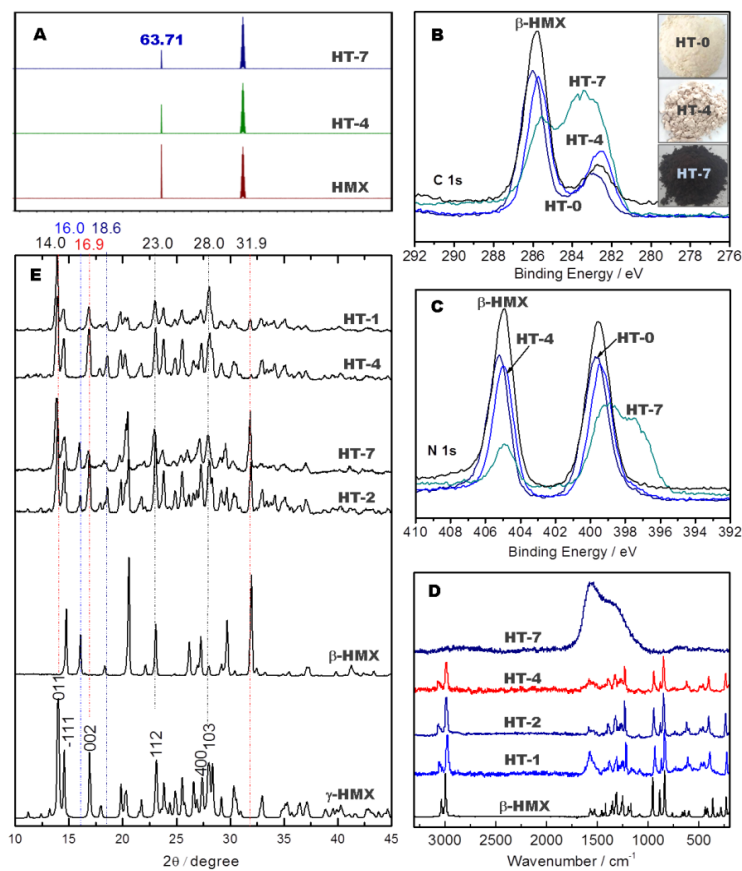

**Fig. 3. The atomic and molecular spectrometry.** (**A**), the $^{13}$C NMR spectra of the starting material HMX, products HT-4 and HT-7. (**B**) and (**C**) C 1s and N 1s XPS spectra of β-HMX, as-prepared HT-0, HT-4 and HT-7 crystalline powders, respectively. (**D**) Raman spectra of β-HMX, and HT-1, HT-2, HT-4 and HT-7 crystals powders, where the black sample H-7 shows a broad peak only due to high absorption coefficient. (**E**) Powder X-Ray spectra of β-HMX, γ-HMX, and asprepared HT-1, HT-4, HT-7 crystals.

It is easy to grow micron-sized doped HMX crystals, but these crystals are not as homogenous as their as-prepared powdered counterparts due to random doping of TAGP layers with different sizes during slow crystallization process, resulting in uncontinuous color (Fig. 4 **A1**-**A3**). In this case, it is extremely difficult to clarify how these TAGP layers being intercalated in the crystals of these modified HMX. Interestingly, we separated the dopants TAGP layers and the HMX molecules using mixture solvent of acetone and ethyl acetate, based on which the crystal of squeezed HMX molecules (HT-6) was obtained and its structure was determined (CCDC 1850523). The unit cell parameters of its crystal lattice are as follows: monoclinic, space group P21/c (No. 14), with $a = 6.527(5)$ Å, $b = 10.926(7)$ Å, $c = 8.699(7)$ Å, $\beta = 124.59(2)°$, $V = 510.6(7)$ Å$^3$, $Z = 2$, $D_{calc} = 1.926$ g cm$^{-3}$. The conformation of squeezed HMX molecules and the resulting HT-6 crystal is very close to that of β-HMX when it was hydrostatically pressed at pressure of 0.2 GPa, named as HMX-II (17). The refined cell parameters of HMX-II are as follows: space group P21/c, $a = 6.495(\pm0.014)$ Å, $b = 10.952(\pm0.010)$ Å, $c = 8.693(\pm0.024)$ Å, $\beta = 124.53(\pm0.20)$, $D_{calc} = 1.931$ g cm$^{-3}$. These pressed/squeezed structures are both in consistent and proportional to that of β-HMX at ambient conditions: a = 6.54 Å, b=11.05 Å, c=8.70 Å, $\beta = 124.30°$, $D_{calc}$ =1.893 g cm$^{-3}$ at ambient conditions (21). The same single crystal X-ray experiment has been done on squeezed RDX, but little change has been observed for its conformation (CCDC 1850522), even though the corresponding doped RDX crystal has a density of 2.04 g cm$^{-3}$. It means that the doping method of TAGP is universal, but the extent of conformation change greatly depends on the rigidity of the molecules, since the HMX molecules with eight-membered ring are much more flexible than RDX. It reveals that after squeezing by the TAGP layers, the conformation of HMX was largely changed (Tables S5-S10). In presence of the solvent molecules (liquid state), such change is irreversible, and hence the new conformation maintained after recrystallization even without the dopants, resulting in a higher packing density.

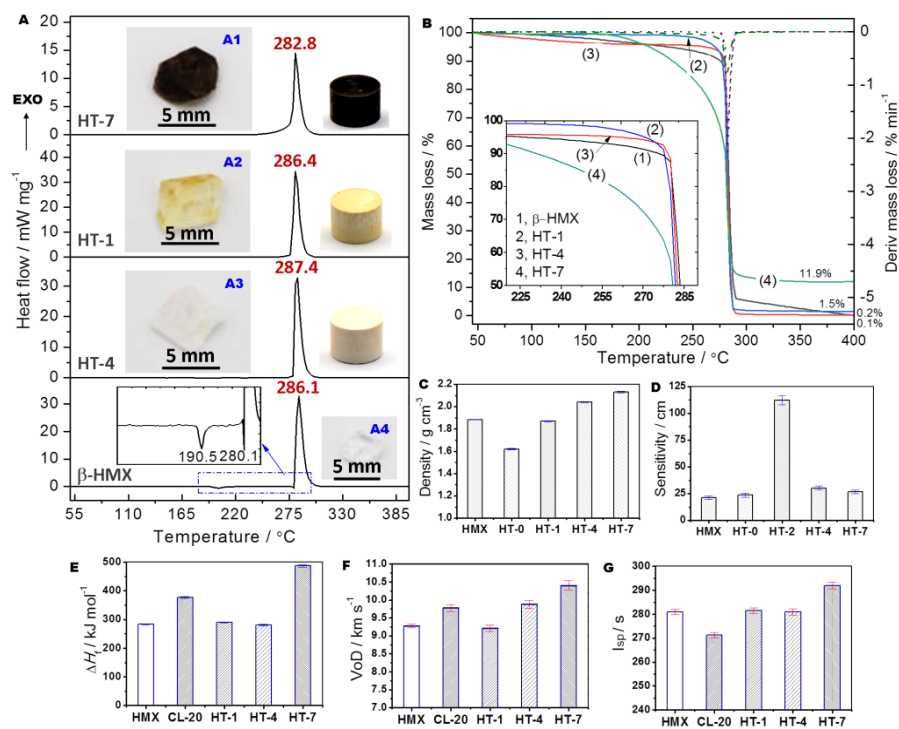

**Fig. 4. Thermal behavior, density, safety and performances.** (**A**) and (**B**), Non-isothermal DSC and TG curves of starting material β-HMX, products HT-1, HT-4 and HT-7 with heating rate of 10 K min$^{-1}$, where the products show no endothermic peaks due to exclusion of polymorphic transition and they have obvious different mass loss behavior due to variation in doping content. (**A1**-**A4**), the digital photos of the selected micron-sized modified HMX crystals obtained by recrystallization of the corresponding powdered HMX with TAGP dopants and pristine β-HMX using the method of naturally slow dissolving of ethanol as the anti-solvent in DMSO solutions of these materials. The corresponding cylindrical of the charges that were used for heat of combustion tests are in a dimension of Φ10 × 10 mm. (**C**), comparison of the crystal density measured by densitometer using helium as the filled gas. (**D**), the impact sensitivity (energy) of raw HMX and the selected modified HMX, evaluated by using a 5 kg dropping hammer according to military standard documented as GJB 772.206-1989. (**E**), (**F**) and (**G**), experimental heat of formation ($\Delta H_f$), calculated detonation velocity (VoD) and specific impulse ($I_{sp}$) of raw β-HMX and the typical TAGP doped HMX crystals under the equal conditions.

Polymorphic HMX molecules have different arrangements in unit cells and thus display different solid-state properties like density, thermodynamic and physicochemical properties including dissolution rate, stability, and hardness, as well as spectroscopic properties. These properties are strongly related to its product quality and performance. As the most stable form at room temperature, β-HMX is stable below the temperature of 185 ℃ (22), after which a polymorphic transition β → δ occurs showing the first endothermic peak at 190.5 ℃ (Fig. 4A) and followed by partial melting with peak temperature of 280.1 ℃ before decomposition. It has been reported that under the pressure over 0.12 GPa, such transition was excluded and HMX decomposes in the state of β-form (17). Interestingly, it is also the case for HT-1, H-4 and HT-7, where no endothermic peak was shown before their decomposition. More importantly, the thermal stability has been improved in the cases of HT-1 and HT-4. It means that after squeezing of TAGP layers as dopants, the HMX molecules with modified conformation and extra constraining force is no longer sensitive to the temperature. In addition, interestingly, there is no heat change in the temperature range of 180-220 ℃ (Fig. 4A, Tables S12 and S14), which is supposed to be the exothermic decomposition of TAGP layers (18). It further indicates successful intercalation of these 2D materials in the crystal of HMX, and therefore the hybrid crystal lattice stabilizes TAGP structure, resulting in a complex single exothermic peak.

In terms of the decomposition heat under the same experimental conditions, the TAGP-doped HMX is higher than of pristine HMX (1574 J g$^{-1}$) except HT-7, and the highest heat produced by HT-2 is about 1688 J g$^{-1}$. More importantly, HT-2 is also the safest materials in all presented samples with an impact sensitivity ($I_m$) of 112 cm (54.9 J, Fig. 4D), whereas it is 24 cm (11.8 J) in the case of pristine HMX. One has to note that these $I_m$ values are obtained based on the Chinese military standard, which is usually over twice larger than that of BAM standard. As shown in Table S14, the densities of HT-4 and HT-7 are around 2.04 and 2.13 g cm$^{-3}$, respectively, even higher than that of currently used the most powerful compound ε-CL-20. In comparison, HT-0 has the lowest density of 1.62 g cm$^{-3}$, which is much lower than that of pristine HMX and even close to that of TAGP material (1.54 g cm$^{-3}$). It means that the intercalation of TAGP under different experimental conditions would result in various states of molecular assembling with different stacking density, which makes the performances of the doped energetic crystals tunable. One can also notice that the decomposition residues of HT-0 and HT-7 (Table S12) are much higher than the other materials, due to nonvolatile products from the higher content of TAGP dopants. It has been reported that the TAGP decomposes with a residue of about 45.4 wt.% (18), and herein the HT-7 contains about 24.67 wt.% TAGP, resulting in a residue of 11.9 wt.% (fig. S15, 45.7%×24.67%=11.2%), which further validates the GPC results (fig. S13).

In order to compare the performances of these new EMs, as a routine work, the heat of formation was calculated based on the experimental heat of combustion data. Then the detonation velocity (VoD) and ground specific impulse ($I_{sp}$) was calculated by using commercial code "Explo-5" based on the experimental data: density, heat of formation ($\Delta H_f$), as well as elemental contents. For better comparisons, the pristine HMX as a starting material was also measured in all cases, resulting in slightly different formula (Table S14). The heat of combustion values ($\Delta H_c$) varies a lot, and the $\Delta H_c$ per unit mass are all increased after doping, but the $\Delta H_f$ of β-HMX, HT-0, HT-1 and HT-4 do not differ too much due to corresponding changes in elemental contents (Fig. 4E). Even by using the experimental formula, the obtained VoD of pristine β-HMX (9277 m s$^{-1}$) is very close to the average value of the reported VoD using its theoretical formula. It has been shown that the heat of combustion ($\Delta H_c$) of the doped HMX is higher than that of pristine HMX, especially in the case of HT-7, whose $\Delta H_c$ is as high as 10.36 kJ g$^{-1}$. As a common fact, the performances of EMs are highly dependent on their density, especially in the case of detonation pressure ($P_{C-J}$). One could notice that the HT-7 reaches the $P_{C-J}$ of 53.87 GPa (Table S14), which is 35% higher than that of the pristine HMX, even higher than currently in-service the most powerful energetic compound ε-CL-20 (44.98 GPa). In terms of VoD and $I_{sp}$, the HT-4 and HT-7 are both higher than that of pristine HMX and even ε-CL-20 (Fig.4 F and G). In particular, the calculated values of VoD for HT-7 and HT-4 are 10.40 and 9.88 km s$^{-1}$, respectively. They would be widely used in both military and civil applications if these values were experimentally confirmed. Also, the $I_{sp}$ of HT-7 reaches about 292 s, making it a very promising propellant component for future space explorations.

In summary, we demonstrated herein a new syntheses route to *in-situ* generation of TAGP in presence of HMX molecules to modify and improve the performance of the resulted hybrid crystalline EMs. It is facile and efficient approach to improve the energy density of the nitramine crystals, resulting in fairly good 2-D materials reinforced crystal structures. More importantly, dual-effect of constraining and doping of TAGP greatly increased the density of resulted crystals, and hence the energetic performances are significantly improved, overpassing the currently used most powerful energetic compound CL-20. We believe that the successful demonstration of this interesting "net-fishing" system in crystal engineering or material science would somehow inspire more scientists to conceive better ideas in construction of novel crystalline materials for various applications.

## REFERENCES

[1] Q. H. Zhang, J. M. Shreeve, Energetic Ionic Liquids as Explosives and Propellant Fuels: A New Journey of Ionic Liquid Chemistry, *Chem Rev*, **114**(20), 10527-10574 (2014).

[2] W. He, P. J. Liu, G. Q. He, M. Gozin, Q. L. Yan, Highly Reactive Metastable Intermixed Composites (MICs): Preparations and Characterizations. *Adv Mater*, doi:10.1002/adma.201706293 (2018).

[3] Y. Wang, Y. Liu, S. Song, Z.J. Yang, X.J. Qi, K.C. Wang, Y. Liu, Q. H. Zhang, Y. Tian, Accelerating the Discovery of Insensitive High Energy-density Materials by a Materials Genome Approach, *Nature Comm*, doi: 10.1038/s41467-018-04897-z (2018).

[4] C. Zhang, C.G. Sun, B.C. Hu, C.M. Yu, M. Lu, Synthesis and Characterization of the Pentazolate Anion Cyclo-$N_5^-$ in $(N_5)_6(H_3O)_3(NH_4)_4Cl$, *Science*, **355**, 374-376 (2017).

[5] Y.G. Xu, Q. Wang, C. Shen, Q.H. Lin, P.C. Wang, M. Lu, A Series of Energetic Metal Pentazolate Hydrates, *Nature*, **549**, 78–81 (2017).

[6] B. Tan, M. Huang, X. Long, J. Li, X. Yuan, R. Xu, From planes to cluster: The Design of Polynitrogen Molecules, *Inter J Quant Chem*, **115** (2), 84-89 (2015).

[7] Y. Li, J. Hao, H. Liu, S. Lu, J.S. Tse, High-energy Density and Super-hard Nitrogen-rich B-N Compounds, *Phys Rev Lett*, **115**(10), art. no. 105502 (2015).

[8] Q.-L. Yan, M. Gozin, F.-Q. Zhao, A. Cohen, S.-P. Pang, High Energetic Compositions Based on Functionalized Carbon Nanomaterials, *Nanoscale* **8**, 4799-4851 (2016).

[9] R.P. Dias, I.F. Silvera, Observation of the Wigner-Huntington Transition to Metallic Hydrogen, *Science*, **355**, 15-718 (2017). doi: 10.1126/science.aal1579.

[10] M. J. Lipp, W. J. Evans, B. J. Baer, C.-S. Yoo, High-energy-density Extended CO solid, *Nature Mater*, **4**, 211-215 (2005).

[11] Q.-L. Yan, F.-Q. Zhao, K.K. Kuo, X.-H. Zhang, S. Zeman, L. T. DeLuca, Catalytic Effects of Nano Additives on Decomposition and Combustion of RDX-, HMX-, and AP-Based Energetic Compositions, *Prog Energ Combust Sci* **57**, 75–136 (2016).

[12] Nandi, A.K., Ghosh, M., Sutar, V.B., Pandey, R.K. Surface Coating of Cyclotetramethylenetetranitramine (HMX) crystals with the insensitive high explosive 1,3,5-triamino-2,4,6-trinitrobenzene (TATB), *Cent Eur J Energet Mater*, **9** (2), 119-130 (2012).

[13] Ø. H. Johansen, J. D. Kristiansen, R. Gjersøe, A. Berg, T. Halvorsen, K.-T. Smith, G. Nevstad, RDX and HMX with Reduced Sensitivity Towards Shock Initiation: RS-RDX and RS-HMX, *Propellants, Explos, Pyrotech*, **33**, 20–24 (2008).

[14] Q. Wang, X. Feng, S. Wang, N. Song, Y. Chen, W. Tong, Y. Han, L. Yang, B. Wang, Metal-Organic Framework Templated Synthesis of Copper Azide as the Primary Explosive with Low Electrostatic Sensitivity and Excellent Initiation Ability, *Adv. Mater.* **28**, 5837 (2016).

[15] O. Bolton, A. J. Matzger, Improved Stability and Smart-Material Functionality Realized in an Energetic Cocrystal, *Angew. Chem. Int. Ed.*, **50**, 8960–8963 (2011).

[16] Q.-L. Yan, P.-J. Liu, A.-F. He, J.-K. Zhang, Y. Ma, H.-X. Hao, F.-Q. Zhao, M. Gozin, Photosensitive but Mechanically Insensitive Graphene Oxide-Carbohydrazide-Metal Hybrid Crystalline Energetic Nanomaterials, *Chem Eng J* **338**, 240–247 (2018). doi.org/10.1016/j.cej.2017.12.140.

[17] C.-S. Yoo, H. Cynn, Equation of state, Phase transition, Decomposition of β-HMX at High Pressures, *J Chem Phys* **111**(22), 10229-10235 (1999). doi:10.1063/1.480341

[18] Q.-L. Yan, A. Cohen, A. K. Chinnam, N. Petrutik, A. Shlomovich, L. Burstein, M. Gozin, Layered 2D Triaminoguanidine-Glyoxal Polymer and Its Transition Metal Complexes, as Novel Insensitive Energetic Nanomaterials. *J Mater Chem A* **4**, 18401-08 (2016).

[19] T. Xu, Y. Xiong, F. Zhong, L. Wang, X. Hao, H. Wang, Chemical Transformation of HMX Induced by Low Energy Electron Beam Irradiation, *Propell. Explos. Pyrot.*, **36** (6), 499-504 (2011).

[20] T. B. Brill, R. J. Karpowicz, Solid Phase Transition Kinetics: The Role of Intermolecular Forces in the Condensed-phase Decomposition of HMX. *J Phys Chem*, **86**, 4260-4265 (1982).

[21] C. S. Choi, H. P. Boutin, *Acta Crystallogr., Sect. B: Struct. Crystallogr. Cryst. Chem.* **26**, 1235 (1970).

[22] M. Herrmann, W. Engel, N. Eisenreich, Thermal Expansion, Transitions, Sensitivities and Burning Rates of HMX, *Propell. Explos. Pyrot.*, **17**, 190-195 (1992).

## ACKNOWLEDGMENTS

This work was supported by "Thousand Youth Talents Plan" with project code of 17GH030127, and it was also partially supported by funding of National Natural Science Foundation of China (51776176). We thank Mr. D.-Y. Tang in our group for analyses of XPS spectra; Dr. Z. Y. Zhang (Shanghai Institute of Materia Medica, Chinese Academy of Sciences) for testing of crystal structure and crystallographic analysis; Additional data are in the supplementary materials. Crystallographic parameters are available free of charge from the Cambridge Crystallographic Data Centre under references CCDC 1850522 and CCDC 1850523.

## SUPPLEMENTARY MATERIALS

https://www.nature.com/content/XXXXX

Materials and Methods
Figs. S1 to S15
Tables S1 to S14
References (S1–S8)